\documentclass[nofootinbib]{revtex4}
\usepackage{graphicx}
\usepackage{dcolumn}
\usepackage{amsmath,amssymb,epsfig}
\usepackage{paralist}
\usepackage{comment}
\usepackage{graphicx}
\usepackage{multirow}
\usepackage{color,soul}
\usepackage{suffix}
\usepackage{mathtools}

\allowdisplaybreaks

\renewcommand{\vec}[1]{\boldsymbol{\mathrm{#1}}}

\begin{document}

\title{Photometric imaging with the solar gravitational lens}

\author{Slava G. Turyshev$^{1}$, Viktor T. Toth$^2$}

\affiliation{\vskip 3pt
$^1$Jet Propulsion Laboratory, California Institute of Technology,\\
4800 Oak Grove Drive, Pasadena, CA 91109-0899, USA}

\affiliation{\vskip 3pt
$^2$Ottawa, Ontario K1N 9H5, Canada}

\date{\today}

\begin{abstract}

We discuss the optical properties of the solar gravitational lens (SGL). We estimate the power of the EM field received by an imaging telescope. Studying the behavior of the EM field at the photometric detector, we develop expressions that describe the received power from a point source as well as from an extended resolved source.  We model the source as a disk with uniform surface brightness and study the contribution of blur to a particular image pixel. To describe this process, we develop expressions describing the power received from the directly imaged region of the exoplanet, from the rest of the exoplanet and also the power for off-image pointing. We study the SGL's  amplification and its angular resolution in the case of observing an extended source with a modest size telescope. The results can be applied to direct imaging of exoplanets using the SGL.

\end{abstract}


\maketitle

\section{Introduction}
\label{sec:aintro}

According to Einstein's general theory of relativity,  the solar gravitational field changes the refractive properties of spacetime. This causes the trajectories of light to bend towards the Sun, resulting in the solar gravitational lens (SGL) \cite{vonEshleman:1979,Turyshev:2017,Turyshev-Toth:2017}. With significant light amplification and impressive angular resolution of up to $2\times 10^{11}$ and $\sim 0.5$~nano-arcseconds, correspondingly (both for $\lambda =1~\mu$m), the SGL may be used to image distant and faint sources.  The impact of the solar corona on the optical properties of the SGL was examined in \cite{Turyshev-Toth:2019-JOO,Turyshev-Toth:2019,Turyshev-Toth:2019-EJP}. This study led to a conclusion that although for radio and microwave frequencies the optical properties of the SGL are severely affected by the plasma in the solar corona, for visible and infrared wavelengths, the contribution by the solar corona is negligible.  The topic of practical use of the SGL for high-resolution imaging and spectroscopy of distant, faint objects recently also received significant attention, leading to the novel concept of a mission capable of reaching the focal region of the SGL and operating there for an extended period of time  \cite{Turyshev-etal:2018,Turyshev-etal:2018-wp}.

Most of the prior work on the SGL dealt with studying the light from point sources at infinite distances from the lens. In \cite{Turyshev-Toth:2019-extend}, we initiated the discussion of using the SGL to image extended sources positioned at large but finite distances from the Sun, thus, moving closer to a possible practical use of this natural phenomenon. While considering the optical properties of the SGL in this case, it was realized that extended sources present an interesting challenge for imaging with the SGL. This challenge relates to a significant blurring of the images, which results in mixing light received from many widely separated areas of the surface of the source and distributing it across the entire image. Several methods of overcoming this challenge were developed. In particular,  \cite{Turyshev-etal:2019-Decadal} demonstrated that the blurring can be removed by relying on modern deconvolution techniques developed to process microlensing images.

In this paper we study the impact of blurring on photometric signals received  when imaging extended resolved sources with a modest size telescope. Our goal is to provide the tools capable of estimating the power of the signal and the contribution of the blur to that signal.
Our paper is organized as follows:
Section~\ref{sec:EM-field} introduces the SGL and the solution for the  electromagnetic (EM) field in the image plane.
Section \ref{sec:image-photom} discusses photometric imaging with the SGL for both point and extended sources.
In Section \ref{sec:ampl-res} we study the realistic amplification of the SGL and its angular resolution in the case of observing extended sources with a realistic telescope.
In Section \ref{sec:disc} we discuss results and explore avenues for the next phase of our investigation of the SGL.

\section{EM field at the interference region}
\label{sec:EM-field}

In \cite{Turyshev-Toth:2019-extend}, we considered light from an extended resolved source positioned at a large but finite distance, $r_0$, from the Sun. We parameterize the problem using a cylindrical coordinate system $(\rho,z,\phi)$, with the $z$-axis corresponding to the preferred axis: a line connecting a preselected (e.g., central) point in the source to the center of the Sun. Furthermore, we characterize points in the image plane and the source plane (both perpendicular to the $z$-axis) using vector coordinates $\vec{x}$ and $\vec{x}'$, respectively.

Considering light, that is to say, a high-frequency EM wave (i.e., neglecting terms $\propto(kr)^{-1}$ where $k=2\pi/\lambda$ is the wavenumber) and for $r\gg r_g$ (where $r_g=2GM_\odot/c^2$ is the Sun's Schwarzschild radius), for a source positioned at a large but finite distance from the Sun, $z_0$, we derived the components of the EM field near the optical axis at heliocentric distances of $z\geq b^2/(2r_g)$, with $b$ being the impact parameter. For a given impact parameter $b$, up to terms of ${\cal O}(\rho^2/z^2, b/z_0)$ in the EM field's amplitude, the components of the EM field take the form
{}
\begin{eqnarray}
    \left( \begin{aligned}
{E}_\rho& \\
{H}_\rho& \\
  \end{aligned} \right) =    \left( \begin{aligned}
{H}_\phi& \\
-{E}_\phi& \\
  \end{aligned} \right)&=&
    \frac{ {E}^{\tt s}_0}{{\overline z}+z_0}
  \sqrt{2\pi kr_g}e^{i\sigma_0}
  J_0\Big(\frac{2\pi}{\lambda}
\sqrt{\frac{2r_g}{\overline z}}
|{\vec x}+\frac{\overline z}{{ z}_0}{\vec x'}|\Big)
    e^{i\big(k(r+r_0+r_g\ln 2k(r+r_0))-\omega t\big)}
 \left( \begin{aligned}
 \cos\phi& \\
 \sin\phi& \\
  \end{aligned} \right),
  \label{eq:DB-sol-rho}
\end{eqnarray}
where the $z$-components of the EM wave behave as $({E}_z, {H}_z)\sim {\cal O}({\rho}/{z},b/z_0)$. The quantity $\overline z=z(1+z/z_0+{\cal O}(z^2/z_0^2))$ denotes heliocentric distances along the line connecting the point source and the center of the Sun. Result (\ref{eq:DB-sol-rho}) is valid for forward scattering when $\theta+b/z_0\approx 0$, or when $0\leq \rho\leq r_g$ \cite{Turyshev-Toth:2019-extend}.

We can describe the imaging of an extended source. For that, we use the solution for the EM field (\ref{eq:DB-sol-rho}) and study the Poynting vector, ${\vec S}=(c/4\pi)\big<\overline{[{\rm Re}{\vec E}\times {\rm Re}{\vec H}]}\big>$, that describes the energy flux at the image plane \cite{Turyshev-Toth:2019-extend}. Normalizing this flux to an empty spacetime (no gravitational lens present), $|{\vec S}_0|=(c/8\pi){E^{\tt s}_0}^2/({\overline z}+z_0)^2$, we define the quantity called the point-spread function (PSF): ${ \mu}_{\tt SGL}=|{\vec S}|/|{\vec S}_0|$. Thus, we use the PSF of the SGL, expressed as a function of the location $\vec{x}$ of a point source at a finite distance from the Sun, and a location $\vec{x'}$ in the image plane:
{}
\begin{eqnarray}
{ \mu}_{\tt SGL}({\vec x},{\vec x}')&=&
\mu_0J^2_0\Big(\frac{2\pi}{\lambda}
\sqrt{\frac{2r_g}{\overline z}}
|{\vec x}+\frac{\overline z}{{ z}_0}{\vec x'}|\Big),
\qquad {\rm with} \qquad
\mu_0=\frac{4\pi^2}{1-e^{-4\pi^2 r_g/\lambda}}
\frac{r_g}{\lambda}.
\label{eq:S_z*6z-mu2}
\end{eqnarray}

As was shown in Fig.~6 of \cite{Turyshev-Toth:2017}, the PSF (\ref{eq:S_z*6z-mu2}) is surprisingly narrow. Depending on the impact parameter, its first zero appears at the distance $\rho_{\tt SGL0}\simeq 4.5\,(\lambda/1~\mu{\rm m})(b/R_\oplus)$ cm from the optical axis (see Eq.~(142) in \cite{Turyshev-Toth:2017}). Thus, to observe the wave optical behavior of the SGL, a very small telescope with aperture $d\propto \rho_{\tt SGL0}$ would be required.  Such a small telescope may not be compatible with a coronagraph, as it cannot resolve the disk of the Sun. Increasing the aperture of the telescope may result in moving outside the wave optics regime. How such an increase affects the optical properties of the SGL is what we investigate.

Equation (\ref{eq:S_z*6z-mu2}) allows us to study the image formation process, develop realistic imaging scenarios, and perform relevant simulations. Considering the distance from the optical axis, $\rho=|{\vec x}+({\overline z}/{{ z}_0}){\vec x'}|$, we see that away from the optical axis, when the argument of the Bessel function $k \sqrt{{2r_g}/{\overline z}}\, \rho\gg 1$ is large, the PSF  (\ref{eq:S_z*6z-mu2}) behaves as
{}
\begin{eqnarray}
{ \mu}_{\tt SGL}&\simeq& \frac{\sqrt{2r_g \overline z}}{\rho}=\frac{b}{\rho}=6.96\times 10^8 \Big(\frac{b}{R_\odot}\Big)\Big(\frac{1~{\rm m}}{\rho}\Big),
\label{eq:psf*}
\end{eqnarray}
where $b=\sqrt{2r_g \overline z}$ is the impact parameter for a given heliocentric distance, $\overline z$, on the optical axis.

Considering (\ref{eq:S_z*6z-mu2}) and (\ref{eq:psf*}) we can make interesting conclusions.  We notice that for a very small telescope that is positioned exactly on the optical axis, $\rho=0$, we may use the expression (\ref{eq:S_z*6z-mu2}) to describe light amplification. This maximal amplification, $\mu_0\simeq 4\pi^2 r_g/\lambda=1.17 \times 10^{11}\,(1\,\mu{\rm m}/\lambda)$, can only be realized with a telescope using a cm-scale aperture. For larger, meter-scale telescopes, the PSF of the SGL may given by (\ref{eq:psf*}), which for the same $\rho$ is identical to (\ref{eq:S_z*6z-mu2}), but is wavelength independent. Therefore, when a telescope with aperture larger than $\sim 30$~cm is used, the PSF of the SGL achromatic.

In addition, we note that in the image plane, away from the optical axis,  the PSF (\ref{eq:psf*}) behaves as $1/\rho$. Such behavior results in significant amplification even when observer is $\rho\sim 10^3$~m away from the optical axis, which is the typical size on an exoplanet image, $r_\oplus$, discussed above. This is in contrast with the PSF of a conventional telescope, given as \cite{Born-Wolf:1999,Turyshev-Toth:2019-extend}
{}
\begin{eqnarray}
\mu_{\tt tel}=\Big(\frac{2J_1(kd\rho/2f)}{kd\rho/2f}\Big)^2
\propto \frac{1}{\rho^3},
\label{eq:psf-norm}
\end{eqnarray}
which falls off much faster with distance from the optical axis, behaving as $\mu_{\tt tel}\propto 1/\rho^3$. Thus, the SGL spreads light much further from the optical axis than the PSF of a conventional telescope. This behavior of the PSF (\ref{eq:S_z*6z-mu2}) and (\ref{eq:psf*}) causes significant blurring that needs to be accounted for in image processing.

Next we address the image formation process in the case of two different types of imaging scenarios. In the first scenario, the observing telescope is used as a photometric detector. In the second case, we investigate the image that forms inside an observing telescope, on its imaging sensor.

\section{Photometric imaging of an extended  source}
\label{sec:image-photom}
\label{sec:SGL-imaging-ext}

In \cite{Turyshev-Toth:2019-extend} we established that, in order to produce an image of an astronomical source, we can assume the source to be noncoherent. Therefore, we need to concern ourselves only with the intensity of light from various points of the source, not its phase. We consider an extended luminous source at a distance of $z_0$ from the Sun, with light, focused by the SGL, detected in an image plane that is at a a distance of $\overline z$ from the Sun. For a source with surface brightness $B(x',y')$ with dimensions of ${\rm W \,m}^{-2}{\rm sr}^{-1}$, the power density, $I_0(x,y)$, received in the image plane is computed by integrating the PSF (\ref{eq:S_z*6z-mu2}) over the surface of the extended source. This can be expressed as
\begin{eqnarray}
I_0(x,y)&=&\frac{\mu_0}{(\overline z+z_0)^2}\iint\displaylimits_{-\infty}^{+\infty}dx'dy'\, B(x',y')\,
J^2_0\Big(\frac{2\pi}{\lambda}\sqrt{\frac{2r_g}{\overline z}}
|{\vec x}+\frac{\overline z}{z_0}{\vec x}'|\Big),
\label{eq:power_dens}
\end{eqnarray}
where $B(x',y')$ is a function with compact support, having nonzero values only within the source's dimensions.

Examining (\ref{eq:S_z*6z-mu2}), we see that the monopole gravitational lens acts as a convex lens by focusing light, according to
{}
\begin{equation}
x=-\frac{\overline z}{z_0}x', \qquad y=-\frac{\overline z}{z_0}y'.
\label{eq:mapping}
\end{equation}
This expression implies that the lens focuses light in the opposite quadrant in the image plane, reducing the image size compared to the source by a factor of ${\overline z}/{z_0}\sim1.0\times 10^{-4}\,({\overline z}/650 ~{\rm AU}) (30~{\rm pc}/z_0)$. The radius of the image of an Earth-like exoplanet at this distance is
{}
\begin{equation}
r_\oplus=\frac{\overline z}{z_0}R_\oplus =1.34 \,\Big(\frac{\overline z}{650 ~{\rm AU}}\Big) \Big(\frac{30~{\rm pc}}{z_0}\Big)~{\rm km}.
\label{eq:r-plus}
\end{equation}

A telescope with aperture $d\ll 2r_\oplus$, centered at a particular point $\vec{x}=(x_0,y_0)$ in the image plane, will receive the signal $P_d(x_0,y_0)=\iint dxdy \,I_0\big(x_0+x,y_0+y\big),$ where the integration is done within the telescope's aperture $|\vec x|\leq d/2$, and with $|{\vec x}_0+{\vec x}|\leq r_\oplus$. This yields a result that depends on the telescope's position on the image plane:
{}
\begin{eqnarray}
P({\vec x}_0)&=&\frac{\mu_0}{(\overline z+z_0)^2}\hskip -5pt\iint\displaylimits_{|{\vec x}|^2\leq (\frac{1}{2}d)^2}\hskip -5pt dxdy \iint\displaylimits_{-\infty}^{+\infty}dx'dy'\, B(x',y')\,
J^2_0\Big(\frac{2\pi}{\lambda}\sqrt{\frac{2r_g}{\overline z}}
|{\vec x}_0+{\vec x}+\frac{\overline z}{z_0}{\vec x}'|\Big).
\label{eq:power_rec2}
\end{eqnarray}

This result represents our starting point for using the SGL for imaging of faint targets positioned at a large but finite distance from the Sun. As $J^2_0(x)$ behaves as $\propto 1/x$, we expect to have nonzero signals at a rather significant distance from the optical axis. We investigate the effect of this behavior on imaging with the SGL.

\subsection{Point source}
\label{sec:point-photo}

Expression (\ref{eq:power_rec2}) is rather complex and, in general, it must be evaluated numerically. However, in the case of a point source, it can be treated analytically. For this, we represent the density of the surface brightness using the Dirac delta function as $B(x',y')=B_{\tt s} \delta(\vec x')$, corresponding to the center of source. With this, we integrate (\ref{eq:power_rec2}) over $d^2{\vec x}'$ to obtain:
{}
\begin{eqnarray}
P_0({\vec x}_0)&=&\frac{\mu_0B_{\tt s}}{({\overline z}+z_0)^2}\hskip -5pt\iint\displaylimits_{|{\vec x}|^2\leq (\frac{1}{2}d)^2}\hskip -5pt dxdy \, J^2_0\Big(\alpha|{\vec x}_0+{\vec x}|\Big),
\qquad{\rm where}\qquad
\alpha=\frac{2\pi}{\lambda}\sqrt{\frac{2r_g}{\overline z}}.
\label{eq:power_rec2i}
\end{eqnarray}

To evaluate (\ref{eq:power_rec2i}), we introduce a polar coordinate system in the image plane, given by $\{\vec x\}=(\rho, \phi)$ and $\{{\vec x}_0\}=(r_0, \phi_0)$ and consider two cases:
\begin{inparaenum}[i)]
\item when the telescope is positioned close to the optical axis, so that $r_0\ll d$, and
\item when the telescope is at a large distance away form the optical axis, $r_0\gg d$.
\end{inparaenum}
This allows us to develop approximate solutions in both cases.

Treating the first case, $r_0\ll d$, we approximate the Bessel function in (\ref{eq:power_rec2i}) with respect to the small parameter $r_0/\rho$, and, keeping only the leading term, we integrate this equation as
{}
\begin{eqnarray}
P_0({\vec x}_0)&=&\frac{\mu_0 B_{\tt s}}{({\overline z}+z_0)^2}\int_0^{2\pi}\hskip -5pt d\phi\int_0^{\frac{d}{2}}\hskip -5pt \rho d\rho J^2_0(\alpha\rho)=\frac{\mu_0 B_{\tt s}}{({\overline z}+z_0)^2}\pi ({\textstyle\frac{1}{2}}d)^2\Big(J^2_0(\alpha{\textstyle\frac{1}{2}}d)+J^2_1(\alpha{\textstyle\frac{1}{2}}d)\Big).
\label{eq:power_rec3i}
\end{eqnarray}

To simplify this expression, we use the approximations for the Bessel functions for large arguments \cite{Abramovitz-Stegun:1965}:
\begin{eqnarray}
J_0(x)\simeq \sqrt{\frac{2}{\pi x}}\cos(x-{\textstyle\frac{\pi}{4}})+{\cal O}\big(x^{-1}\big)
\qquad {\rm and} \qquad
J_1(x)\simeq \sqrt{\frac{2}{\pi x}}\sin(x-{\textstyle\frac{\pi}{4}})+{\cal O}\big(x^{-1}\big),
\label{eq:BF}
\end{eqnarray}
and, taking into account expressions for $\mu_0$ from (\ref{eq:S_z*6z-mu2}), for $\alpha$ from (\ref{eq:power_rec2i}), and also accounting for the fact that ${\overline z}/z_0\ll1$, we express (\ref{eq:power_rec3i}) as
{}
\begin{eqnarray}
P_0(r_0)&=&\pi B_{\tt s} \Big(\frac{d}{z_0}\Big)^2\frac{\sqrt{2r_g \overline z}}{d},
\qquad{\rm for}\qquad
r_0\ll d.
\label{eq:power_rec4i}
\end{eqnarray}

This expression shows that the power received by a telescope from a point (i.e., unresolved) source is insensitive to small deviations from the optical axis in the image plane.

In a similar manner, we consider the case when the telescope is positioned at a large distance from the optical axis, $r_0\gg d$. By expanding the Bessel function in (\ref{eq:power_rec2i}) in terms of the small parameter $\rho/r_0$ and keeping only the leading term, we obtain the following result:
{}
\begin{eqnarray}
P_0({\vec x}_0)&=&\frac{\mu_0 B_{\tt s}}{({\overline z}+z_0)^2}\int_0^{2\pi}\hskip -5pt d\phi\int_0^{\frac{d}{2}}\hskip -5pt \rho d\rho J^2_0(\alpha r_0)=
\frac{\mu_0 B_{\tt s}}{({\overline z}+z_0)^2}\pi ({\textstyle\frac{1}{2}}d)^2
J^2_0(\alpha r_0).
\label{eq:power_rec73i}
\end{eqnarray}
Using the same approximation for the Bessel functions (\ref{eq:BF}) and expressions for $\mu_0$ and $\alpha$, given by (\ref{eq:S_z*6z-mu2}) and (\ref{eq:power_rec2i}), correspondingly, and also by accounting for the fact that ${\overline z}/z_0\ll1$, we present (\ref{eq:power_rec73i}) as
{}
\begin{eqnarray}
P_0(r_0)&=&\pi B_{\tt s}\Big(\frac{d}{2z_0}\Big)^2\frac{\sqrt{2r_g\overline z}}{r_0} \Big(1+\sin\Big[2kr_0\sqrt{\frac{2r_g}{\overline z}}\Big]\Big), \qquad{\rm for}\qquad
r_0\gg d.
\label{eq:power_rec5i}
\end{eqnarray}
This expression clearly exhibits the oscillatory behavior that diminishes as $1/r_0$ with distance from the optical axis. Such behavior is consistent with that expected from the PSF  of the SGL given by (\ref{eq:S_z*6z-mu2}).

Results (\ref{eq:power_rec4i}) and (\ref{eq:power_rec5i})  may be used to estimate the power of the signal received from a distant, unresolved source.

\subsection{Extended source}
\label{sec:extend-photo}
\label{sec:extend-source}

Now we consider the process of imaging an extended, resolved source.
In the most widely considered, practical scenario, a kilometer-scale image plane is sampled by a meter-scale telescope that, while it has the resolution required to employ a coronagraph, is otherwise used as a ``light bucket'', collecting light from the exoplanet.

\begin{figure}
\includegraphics[width=0.39\linewidth]{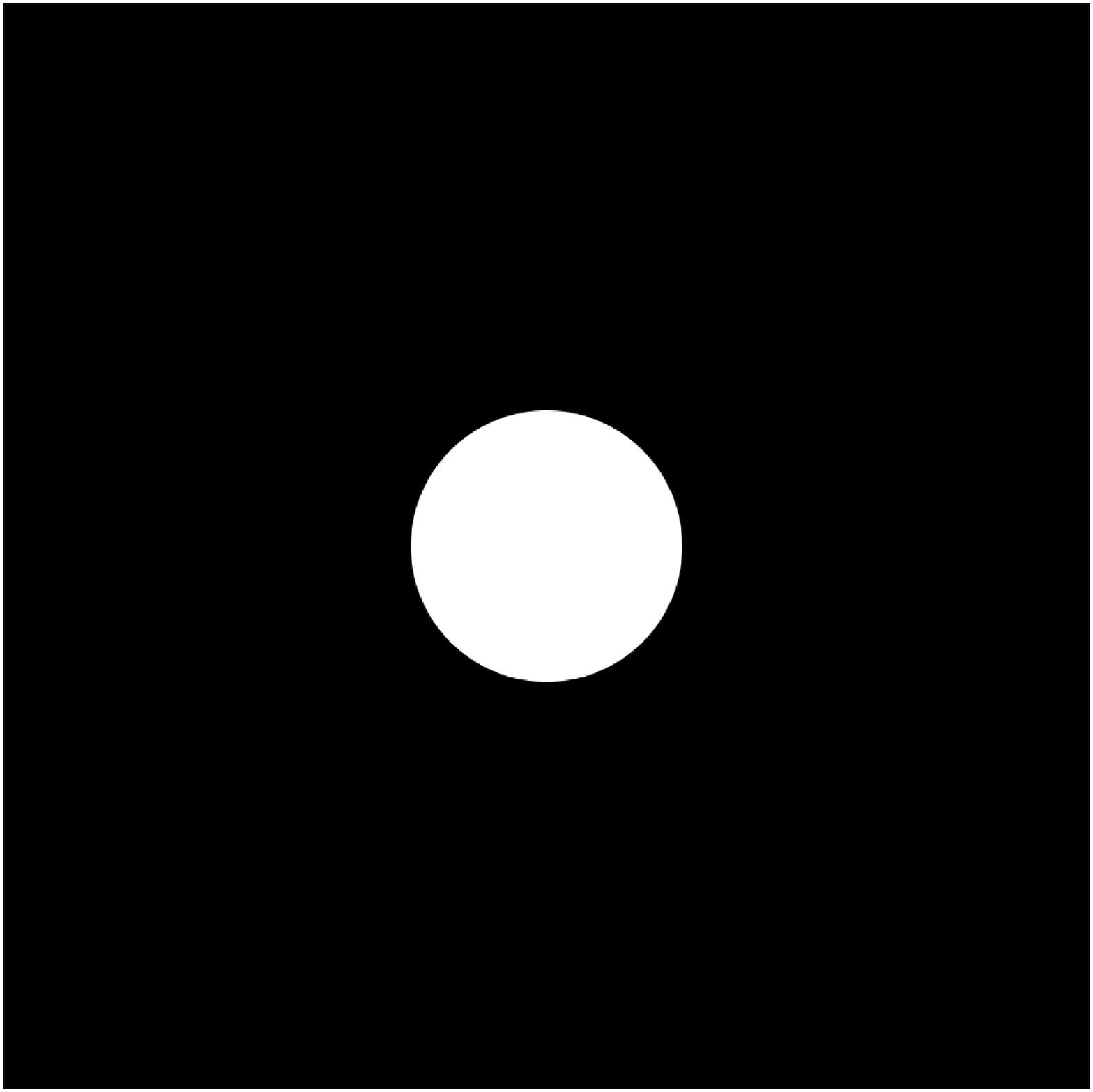}~
\includegraphics[width=0.39\linewidth]{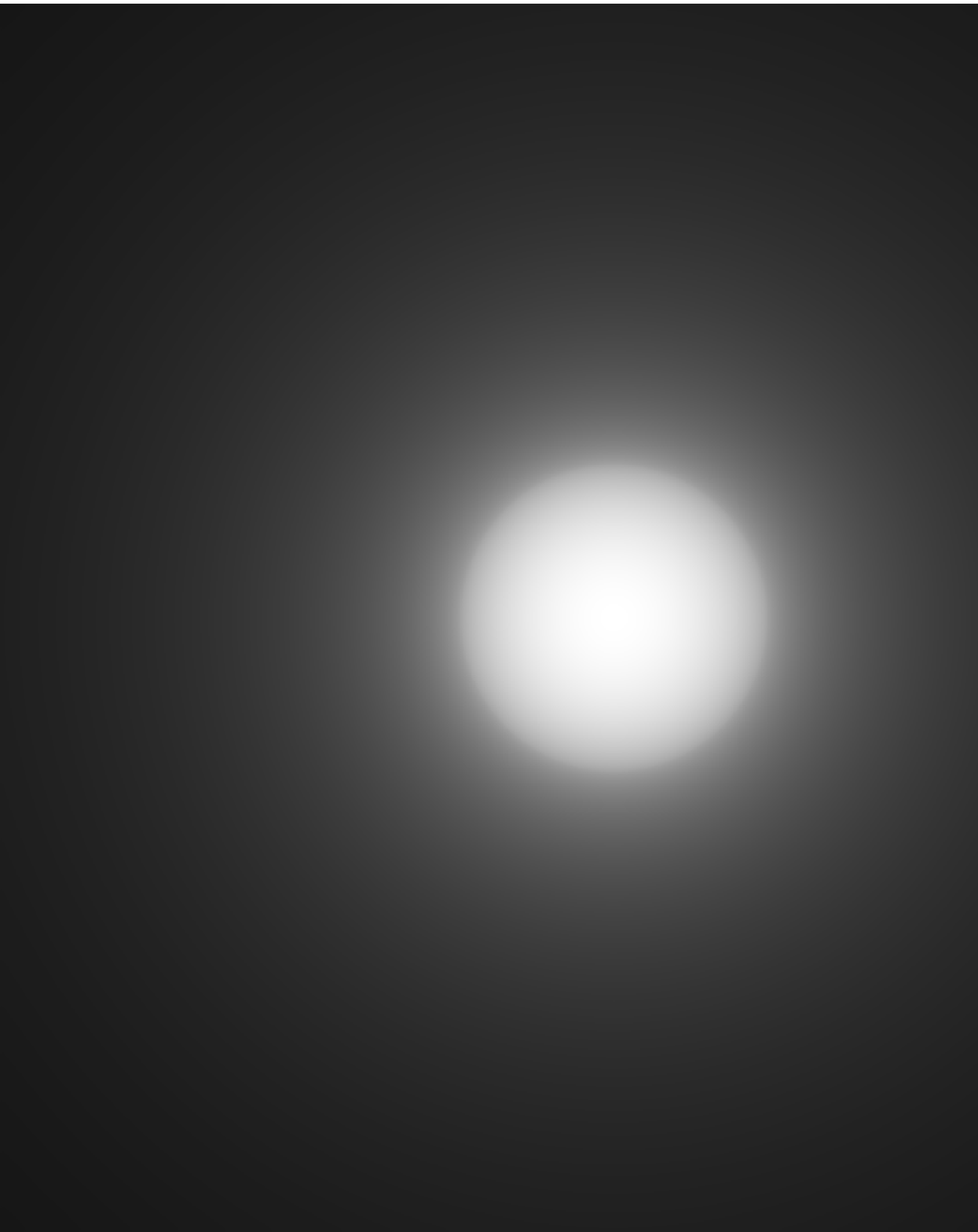}
\caption{\label{fig:blur} Convolving the source with the SGL. Left: a disk of uniform surface brightness. Right: the image of that disk on the image plane of the SGL. The blur is evident both within the image and outside of it.}
\end{figure}

With its PSF (\ref{eq:S_z*6z-mu2})--(\ref{eq:psf*}), the SGL is not a lens with ideal optical properties. This is demonstrated in Fig.~\ref{fig:blur}, which shows the result of a numerical integration of (\ref{eq:power_rec2}) that was achieved by convolving a luminous disk of uniform surface brightness and the PSF of the SGL. Evident in the image is the substantial blurring introduced by the SGL.

\begin{figure}
\includegraphics[scale=0.30]{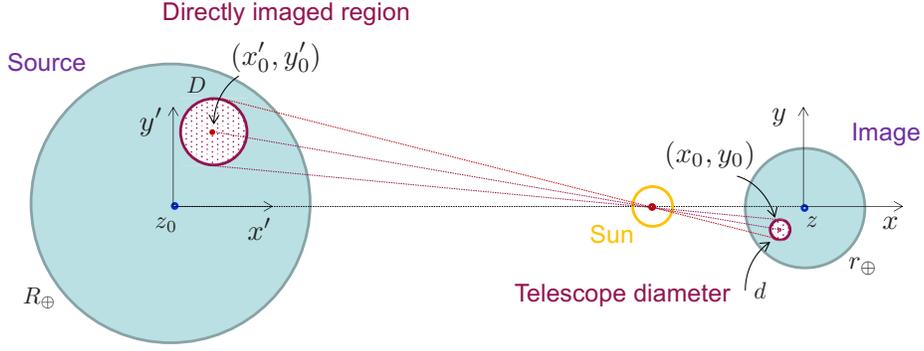}
\caption{\label{fig:regions}Imaging of extended resolved sources with the SGL. The concept of a directly imaged region.}
\end{figure}

To model the behavior seen in Fig.~\ref{fig:blur}, we recognize that the telescope's aperture is much smaller than the image size, $d\ll 2r_\oplus$. This leads us to separate the received signal in two parts: the signal received from the directly imaged region that corresponds to the telescope location in the image plane, and the blur from light received from the rest of the planet.
Based on the SGL's mapping (\ref{eq:power_rec2}) for a given point $(x_0,y_0)$ in the image plane (see Fig.~\ref{fig:regions}), the directly imaged region is in the vicinity of the point $(x'_0,y_0')=-(z_0/\overline z)(x_0,y_0)$ in the source plane.  Furthermore, given the telescope aperture $d$, the directly imaged region in the source plane has the diameter
\begin{equation}
D=\frac{z_0}{\overline z}d,
\label{eq:Dd}
\end{equation}
centered at $(x'_0,y_0')$. The signal that is received from the exoplanet from areas outside this area is causing the blur.

In the general case, the function $B(x',y')$ in Eq.~(\ref{eq:power_dens}) is not known in advance. It characterizes the surface of the exoplanet that is being observed. Numerical methods of image reconstruction and deconvolution must be used to reconstruct $B(x',y')$ from observed values of $P_d(x_0,y_0)$ as part of any observational campaign.

To establish limits on the sensitivity of the SGL and minimum criteria for an observing instrument, however, it is instructive to attempt to evaluate the integral (\ref{eq:power_rec2}) analytically in simple model scenarios. To do that, it is convenient to express the position of the  telescope in the image plane $\{{\vec x}_0\}=(x_0,y_0)$ via its counterpart at the source, $\{{\vec x}_0'\}=(x'_0,y_0')$; see Fig.~\ref{fig:regions} for details.  Using the mapping (\ref{eq:mapping}), we can write:
{}
\begin{equation}
{\vec x}_0=-\frac{\overline z}{z_0}{\vec x}_0'.
\label{eq:mapping*}
\end{equation}
As a result, (\ref{eq:power_rec2}) takes the following form:
{}
\begin{eqnarray}
P({\vec x}_0)&=&\frac{\mu_0}{(\overline z+z_0)^2}\hskip -5pt\iint\displaylimits_{|{\vec x}|^2\leq (\frac{1}{2}d)^2}\hskip -5pt dxdy \iint\displaylimits_{-\infty}^{+\infty}dx'dy'\, B(x',y')\,
J^2_0\Big(\alpha |{\vec x}+\eta({\vec x}'-{\vec x}'_0)|\Big),
\qquad{\rm where}\qquad
\eta=\frac{\overline z}{z_0},
\label{eq:power_rec2*}
\end{eqnarray}
where the surface brightness density, $B(x',y')$, is assumed to be a function with compact support with nonzero values only within the dimensions of the source, and where $\alpha$ is defined in (\ref{eq:power_rec2i}).

We consider the case when the telescope is positioned in the image plane at the distance $0\leq r_0\leq r_\oplus$ from the center of the image. To appreciate the blur contribution, we present the integral over the source as a sum:
{}
\begin{eqnarray}
P_{\tt exo}({\vec x}_0)&=&P_{\tt dir}({\vec x}_0)+P_{\tt blur}({\vec x}_0),
\label{eq:dir+blur}
\end{eqnarray}
where $P_{\tt dir}({\vec x}_0)$ is the power received from the directly imaged region on the source (i.e., for $|{\vec x}'-{\vec x}'_0|\in [0,{\textstyle\frac{1}{2}}D]$) and $P_{\tt blur}({\vec x}_0)$ is the power received by the telescope  from the rest of the planet (i.e., for $|{\vec x}'-{\vec x}'_0|\in [{\textstyle\frac{1}{2}}D,\rho_\oplus(\phi')]$), where $\rho_\oplus(\phi')$ is the radial coordinate of the edge of the source, as seen from the center of the directly imaged region. In addition, as seen in Fig.~\ref{fig:blur}, even a telescope positioned outside the area in the image plane that corresponds to a direct image of the exoplanet receives significant light. This case is also discussed below.

\subsubsection{Power from the directly imaged region}
\label{sec:dir-image-reg}

We begin with estimating the power received from the directly imaged region, $P_{\tt dir}({\vec x}_0)$, which is given by (\ref{eq:power_rec2*})
\begin{eqnarray}
P_{\tt dir}({\vec x}_0)&=&\frac{\mu_0}{(\overline z+z_0)^2}
\int_0^{2\pi}\hskip -5pt d\phi\int_0^{\frac{d}{2}}\hskip -5pt \rho d\rho
\hskip -15pt
\iint\displaylimits_{|{\vec x}'-{\vec x}'_0|\in [0,{\textstyle\frac{1}{2}}D]}
\hskip -15pt dx'dy'\, B(x',y')\,
J^2_0\Big(\alpha \big|{\vec x}+\eta({\vec x}'-{\vec x}'_0)\big|\Big),
\label{eq:power-dir}
\end{eqnarray}
where the integration of over the source is done within the size of the directly imaged region, namely  $|{\vec x}'-{\vec x}'_0|\in [0,{\textstyle\frac{1}{2}}D]$.

To evaluate (\ref{eq:power-dir}), we introduce a polar coordinate system in the image plane, given by $\{\vec x\}=(\rho, \phi)$ and $\{({\vec x}'-{\vec x}_0)\}=(r', \phi')$. Next, we recognize that within the directly imaged region $\eta r' \leq \rho$. Therefore, we may express the Bessel function in terms of the small parameter $ \eta |{\vec x}'-{\vec x}'_0|/|{\vec x}|\equiv \eta r'/\rho\ll 1$, again keeping only the leading term. Remembering $\alpha$ from (\ref{eq:power_rec2i}) and assuming that the density of the source brightness within this region is uniform, $B(x',y')=B_s$, allows to integrate (\ref{eq:power-dir}), thus expressing the power received from the directly imaged region as
\begin{eqnarray}
P_{\tt dir}({\vec x}_0)&=&\frac{\mu_0}{z^2_0}
\int_0^{2\pi}\hskip -5pt d\phi\int_0^{\frac{d}{2}}\hskip -5pt \rho d\rho
\int_0^{2\pi}\hskip -5pt d\phi'\int_0^{\frac{D}{2}}\hskip -5pt r' dr'
B(\rho',\phi')\,J^2_0\big(\alpha\rho\big)=
\frac{\mu_0}{z^2_0}\pi ({\textstyle\frac{1}{2}}d)^2 B_{\tt s}\pi ({\textstyle\frac{1}{2}}D)^2
\Big(
J^2_0\big(\alpha{\textstyle\frac{1}{2}}d\big)+J^2_1\big(\alpha{\textstyle\frac{1}{2}}d\big)\Big).~~~~~
\label{eq:power-dir*19}
\end{eqnarray}

Using the approximations for the Bessel functions (\ref{eq:BF}), taking into account expressions for $\mu_0$ from (\ref{eq:S_z*6z-mu2}), for $D$ from (\ref{eq:Dd}), and for $\alpha$ from (\ref{eq:power_rec2i}), we can express (\ref{eq:power-dir*19}) as
\begin{eqnarray}
P_{\tt dir}(r_0)&=&B_s   \frac{\pi^2 d^3}{4\overline z}\sqrt{\frac{2r_g}{\overline z}}.
\label{eq:Pdir*}
\end{eqnarray}

The power received from the directly imaged region does not depend on the distance to the exoplanet or position of the telescope in the image plane (as we assumed that the planet has uniform brightness), but it strongly depends on the telescope aperture, $d$, and also on the heliocentric distance to the image plane, ${\overline z}$.  These properties are consistent with the imaging of unresolved sources \cite{Born-Wolf:1999}.

\subsubsection{Power from the rest of the planet}
\label{sec:dir-image-blur}

The power of the blur from the rest of the planet, $P_{\tt blur}({\vec x}_0)$, is given by (\ref{eq:power_rec2*}), where the integration in the source plane is done over the rest of the planet that falls outside the directly imaged region. This is a much larger part of the planet within the boundary $|{\vec x}'-{\vec x}'_0|\in [{\textstyle\frac{1}{2}}D,\rho_\oplus(\phi')]$, where $\rho_\oplus(\phi')$ is the radial coordinate of the edge of the source, as seen from the center of the directly imaged region:
{}
\begin{eqnarray}
P_{\tt blur}({\vec x}_0)&=&\frac{\mu_0}{(\overline z+z_0)^2}\int_0^{2\pi}\hskip -5pt d\phi\int_0^{\frac{d}{2}}\hskip -5pt \rho d\rho
\hskip -20pt
\iint\displaylimits_{|{\vec x}'-{\vec x}'_0|\in [{\textstyle\frac{1}{2}}D,\rho_\oplus(\phi')]}
\hskip -25pt dx'dy'\, B(x',y')\,
J^2_0\Big(
\alpha \big|{\vec x}+\eta({\vec x}'-{\vec x}'_0)\big|\Big).
\label{eq:power-blur}
\end{eqnarray}

To evaluate the blur component (\ref{eq:power-blur}), we recognize that in most of the area outside the directly imaged region the following equality is valid: $|{\vec x}|\ll \eta |{\vec x}'-{\vec x}'_0|$. With this, we can expand the Bessel function in (\ref{eq:power-blur}) in terms of the small parameter $|{\vec x}|/(\eta |{\vec x}'-{\vec x}'_0|)\equiv \rho/(\eta r')\ll1$. Keeping only the leading term, we integrate (\ref{eq:power-blur}) over the image plane as
{}
\begin{eqnarray}
P_{\tt blur}({\vec x}_0)&=&\frac{\mu_0}{z^2_0}
\int_0^{2\pi}\hskip -5pt d\phi\int_0^{\frac{d}{2}}\hskip -5pt \rho d\rho
\hskip -20pt
\iint\displaylimits_{|{\vec x}'-{\vec x}'_0|\in [{\textstyle\frac{1}{2}}D,\rho_\oplus(\phi')]}
\hskip -25pt dx'dy'\, B(x',y')\,
J^2_0\big(\alpha\eta r'\big)=
\frac{\mu_0}{z^2_0}\pi ({\textstyle\frac{1}{2}}d)^2
\hskip -30pt
\iint\displaylimits_{|{\vec x}'-{\vec x}'_0|\in [{\textstyle\frac{1}{2}}D,\rho_\oplus(\phi')]}
\hskip -25pt dx'dy'\, B(x',y')\,
 J^2_0\big(\alpha\eta r'\big).~~~~~
\label{eq:power-blur2}
\end{eqnarray}

Next, we introduce a new coordinate system on the source plane, ${\vec x}''$, with the origin at the center of the directly imaged region: ${\vec x}'-{\vec x}'_0={\vec x}''$. As the vector ${\vec x}'_0$ is constant, $dx'dy'=dx''dy''$. Next, in the new coordinate system, we use polar coordinates $(x'',y'')\rightarrow (r'',\phi'')$. We can see that, in these coordinates, the circular edge of the source, $R_\oplus$, is a curve, $\rho_\oplus(\phi'')$, the radial distance of which is given by the following relation:
{}
\begin{eqnarray}
\rho_\oplus(\phi'')
&=&\sqrt{R_\oplus^2-{r'_0}^2\sin^2\phi''}-r'_0\cos\phi''.
\label{eq:rho+}
\end{eqnarray}
With this, and assuming that the source in this region may be characterized by a uniform surface brightness density, $B(x',y')=B_s$, we evaluate (\ref{eq:power-blur2}) as
{}
\begin{eqnarray}
P_{\tt blur}({\vec x}_0)
&=&\frac{\mu_0}{z_0^2}\pi ({\textstyle\frac{1}{2}}d)^2 B_s\int_0^{2\pi} \hskip -3pt d\phi''
\int_{\frac{D}{2}}^{\rho_\oplus}\hskip -3pt r'' dr''
J^2_0\big(\alpha\eta r''\big)=\nonumber\\
&=&\frac{\mu_0}{z_0^2}\pi ({\textstyle\frac{1}{2}}d)^2 B_s\int_0^{2\pi} \hskip -3pt d\phi''\Big\{
\frac{\rho_\oplus^2}{2}\Big(J^2_0\big(\alpha\eta\rho_\oplus\big)+J^2_1\big(\alpha\eta\rho_\oplus\big)\Big)-
\frac{D^2}{8}\Big(J^2_0\big(\alpha\eta{\textstyle\frac{1}{2}}D\big)+J^2_1\big(\alpha\eta{\textstyle\frac{1}{2}}D\big)\Big)\Big\},
\label{eq:power-blur5}
\end{eqnarray}
where $\rho_\oplus=\rho_\oplus(\phi'')$ and $D$ as given by (\ref{eq:rho+}) and (\ref{eq:Dd}), correspondingly.

To simplify the integration of (\ref{eq:power-blur5}),  we use approximations of the Bessel functions (\ref{eq:BF}), which are valid in this region. This allows us to express the integrand as
{}
\begin{eqnarray}
\frac{\rho_\oplus^2}{2}\Big(J^2_0\big(\alpha\eta\rho_\oplus\big)+J^2_1\big(\alpha\eta\rho_\oplus\big)\Big)-
\frac{D^2}{8}\Big(J^2_0\big(\alpha\eta{\textstyle\frac{1}{2}}D\big)+J^2_1\big(\alpha\eta{\textstyle\frac{1}{2}}D\big)\Big)&=&\frac{\rho_\oplus(\phi'')-{\textstyle\frac{1}{2}}D}{\pi\alpha\eta}.
\label{eq:power-int}
\end{eqnarray}

With these results, (\ref{eq:power-blur5}) is evaluated as
{}
\begin{eqnarray}
P_{\tt blur}({\vec x}_0)
&=&B_s \pi ({\textstyle\frac{1}{2}}d)^2  \frac{\mu_0}{z_0^2}
\frac{D}{2\pi\alpha\eta}
\int_0^{2\pi} \hskip -3pt d\phi''\Big(\frac{2\rho_\oplus(\phi'')}{D}-1\Big).
\label{eq:power-blur6*}
\end{eqnarray}

\begin{figure}[t!]
 \vspace{10pt}
  \begin{center}
  \rotatebox{90}{\hskip 35pt {\footnotesize $\epsilon(r_0)$}}
\includegraphics[scale=0.61]{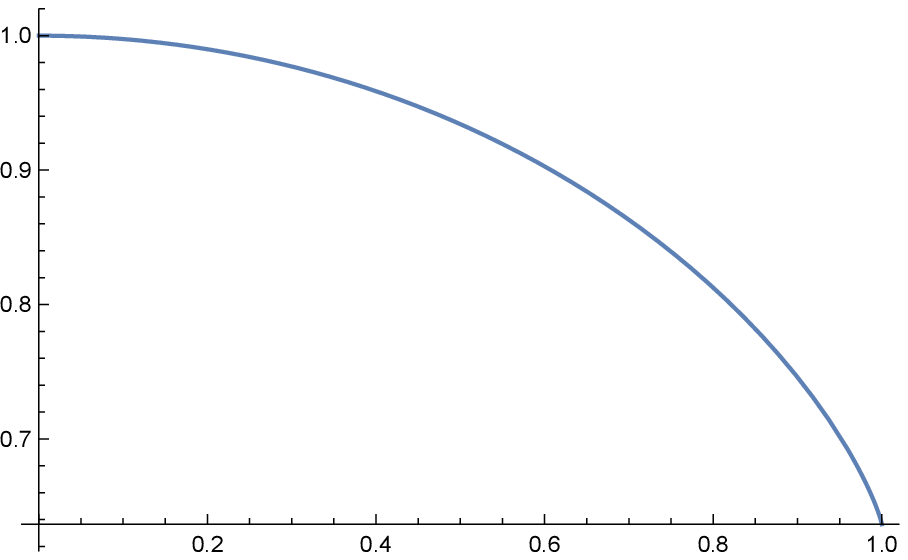}
\hskip -3pt
\rotatebox{90}{\hskip 35pt {\footnotesize  $\beta(r_0)$}}
\includegraphics[scale=0.61]{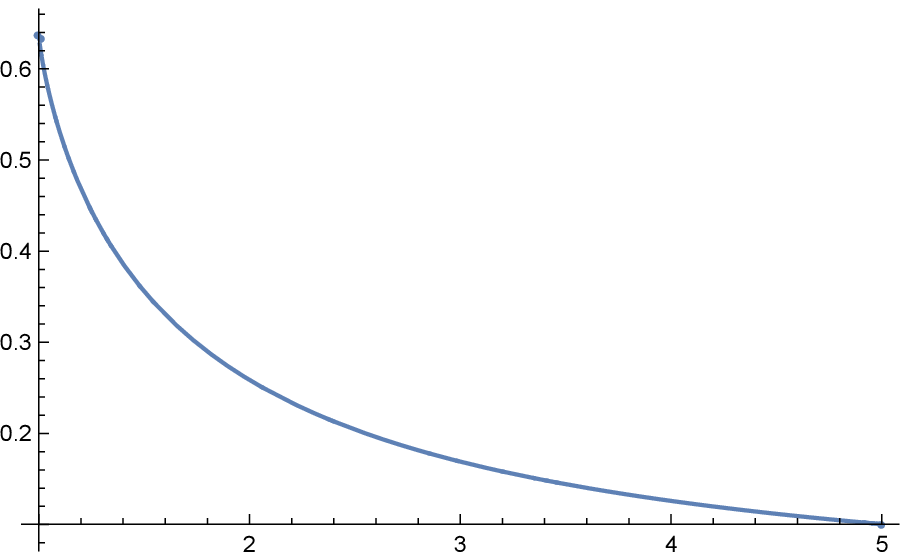}
\hskip -3pt
\rotatebox{90}{\hskip 25pt {\footnotesize $\epsilon(r_0)+\beta(r_0)$}}
\includegraphics[scale=0.61]{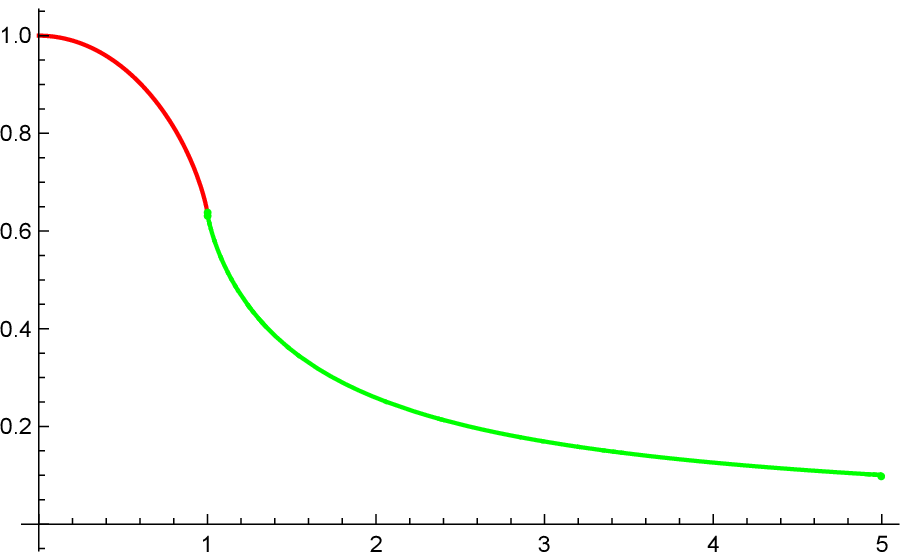}
  \end{center}
  \vspace{-10pt}
  \caption{Left: Behavior of the factor $\epsilon(r_0)$, plotted as a function of $r_0/r_\oplus$. Center: Behavior of the factor $\beta(r_0)$, also as a function of $r_0/r_\oplus$. Right: Combined behavior of the factors $\epsilon(r_0)$ and $\beta(r_0)$ in the relevant region inside and around the image.}
\label{fig:factors}
  \vspace{-5pt}
\end{figure}

Expression (\ref{eq:power-blur6*}) indicates that for a source with uniform surface brightness, there is no azimuthal dependence of blur on the telescope's position within the image and the received power depends only on the separation from the optical axis.
Taking into account expressions for $\mu_0$ from (\ref{eq:S_z*6z-mu2}), for $D$ from (\ref{eq:Dd}), for $\eta$ from (\ref{eq:power_rec2*}), and for $\alpha$ from (\ref{eq:power_rec2i}), we can express (\ref{eq:power-blur6*}) as
{}
\begin{eqnarray}
P_{\tt blur}(r_0)
&=&B_s  \frac{\pi^2 d^3}{4\overline z}\sqrt{\frac{2r_g}{\overline z}}
\Big(\frac{2R_\oplus}{d}\frac{\overline z}{z_0}\epsilon(r_0)-1\Big),
\label{eq:power-blur6*7}
\end{eqnarray}
where the factor $\epsilon(r_0)$  is given by the following expression:
{}
\begin{eqnarray}
\epsilon(r_0)
&=&\frac{1}{2\pi}\int_0^{2\pi} \hskip -3pt d\phi''\sqrt{1-\Big(\frac{r_0}{r_\oplus}\Big)^2\sin^2\phi''}=\frac{2}{\pi}{\tt E}\Big[\Big(\frac{r_0}{r_\oplus}\Big)^2\Big],
\label{eq:eps_r0}
\end{eqnarray}
where ${\tt E}[x]$ is the elliptic integral \cite{Abramovitz-Stegun:1965}.
The behavior of $\epsilon(r_0)$ for the values of $r_0\in[0,r_\oplus]$ is shown in the left-side plot in Fig.~\ref{fig:factors}.

As a result, expression (\ref{eq:power-blur6*7}) may be given as
{}
\begin{eqnarray}
P_{\tt blur}(r_0)
&=&B_s \frac{\pi^2 d^3}{4\overline z}\sqrt{\frac{2r_g}{\overline z}}
\Big(\frac{2R_\oplus}{d}\frac{\overline z}{z_0}\epsilon(r_0)-1\Big).
\label{eq:Pblur*}
\end{eqnarray}
We can see that for a telescope with modest aperture size, $d\ll 2r_\oplus=2R_\oplus (\overline z/z_0)$, the blur contribution is much larger than the power received from the directly imaged region:
{}
\begin{eqnarray}
P_{\tt blur}(r_0)&=&P_{\tt dir}(r_0)\Big(\frac{2R_\oplus}{d}\frac{\overline z}{z_0}\epsilon(r_0)-1\Big).
\label{eq:Pblur*2}
\end{eqnarray}

The total signal received from the exoplanet is computed by summing up the two contributions  as given by  (\ref{eq:dir+blur}).  Then, using (\ref{eq:Pdir*}) and (\ref{eq:Pblur*}) we have, to ${\cal O}\big({\overline z}{/z_0}\big)$:
{}
\begin{eqnarray}
P_{\tt exo}(r_0)&=&P_{\tt dir}(r_0)+P_{\tt blur}(r_0)=B_s\pi^2 d^2  \frac{R_\oplus}{2z_0}\sqrt{\frac{2r_g}{\overline z}}\,\epsilon(r_0), \qquad 0\leq r_0\leq r_\oplus.
\label{eq:Pexo}
\end{eqnarray}

Expression (\ref{eq:Pexo}) is our main result for the case of photometric imaging.  It shows that at every pixel, the signal for the directly imaged region is overwhelmed by the blur. Also, for various pixels, the amount of blur is different. It is highest for the central region of the image and is about  $2/\pi\approx 0.64$ times smaller when considering pixels close to the edge of the image. As we studied the case when the source has a uniform surface brightness, the result (\ref{eq:Pexo}) is independent of the azimuthal angle and is a function only of the distance from the optical axis, $r_0$.

For imaging purposes, we shall treat the power received from the directly imaged region as the signal of interest and the blur from the rest of the planet is the nuisance to signal recovery. Given the SGL's significant light amplification, there is a sufficient signal-to-noise ratio that allows for signal extraction from the blurred data. This can be done by relying on modern deconvolution algorithms \cite{Turyshev-etal:2018}, capable of image recovery under such conditions.

\subsubsection{Blur for an off-image telescope position}
\label{sec:extend-photo-vic}

As we see from Fig.~\ref{fig:blur}, the blurring of the images obtained with the SGL is evident even outside the direct image of an exoplanet.
Therefore, even a telescope positioned at $r_0\geq r_\oplus$ will receive light from the source. In this case, the blur for the off-image position, $P_{\tt off}({\vec x}_0)$, is obtained by integrating over the surface of the source as it is seen from an off-image coordinate system:
{}
\begin{eqnarray}
P_{\tt off}({\vec x}_0)&=&\frac{\mu_0}{(\overline z+z_0)^2}\int_0^{2\pi}\hskip -5pt d\phi\int_0^{\frac{d}{2}}\hskip -5pt \rho d\rho
\hskip -10pt
\iint\displaylimits_{|{\vec x}'-{\vec x}'_0|\leq R_\oplus}
\hskip -10pt dx'dy'\, B(x',y')\,
J^2_0\Big(
\alpha \big|{\vec x}+\eta({\vec x}'-{\vec x}'_0)\big|\Big).
\label{eq:power-*blur}
\end{eqnarray}

Expression (\ref{eq:power-*blur}) gives the power of the light received when telescope is located off-image. The same conditions to derive (\ref{eq:power-blur2}) are valid, so the power received by the telescope takes the same form.
The only difference comes from the fact that we are outside the image, thus, the integration limits change. First, we note that the circular edge of the source, $R_\oplus$, is given by a curve, $\rho_\oplus(\phi'')$, the radial distance of which in this polar coordinate system is given as
\begin{eqnarray}
\rho_\oplus(\phi'') &=&\pm \sqrt{R_\oplus^2-{r'_0}^2\sin^2\phi''}+r'_0\cos\phi'',
\label{eq:rho++}
\end{eqnarray}
with the angle $\phi''$ in this case is defined so that $\phi''=0$ when pointing at the center of the source. The angle $\phi''$ varies only within the range $\phi''\in [\phi_-,\phi_+]$, with $\phi_\pm=\arcsin ({R_\oplus}/{r'_0})$. Given the sign in front of the square root in (\ref{eq:rho++}), for any angle $\phi''$ there will be two solutions for $\rho_\oplus(\phi'')$, given as $(\rho^-_\oplus,\rho^+_\oplus)$.

Assuming that the brightness of the source in this region is uniform, $B(x',y')=B_s$, we use (\ref{eq:power-blur5}) and evaluate (\ref{eq:power-*blur}) for this set of conditions:
{}
\begin{eqnarray}
P_{\tt off}({\vec x}_0)
&=&\frac{\mu_0}{z_0^2}\pi ({\textstyle\frac{1}{2}}d)^2 B_s
\int_{\phi_-}^{\phi_+} \hskip -3pt d\phi''
\int_{\rho^-_\oplus}^{\rho^+_\oplus}\hskip -3pt r'' dr''
J^2_0\big(\alpha\eta r''\big)
=\nonumber\\
&=&\frac{\mu_0}{z_0^2}\pi ({\textstyle\frac{1}{2}}d)^2 B_s
\int_{\phi_-}^{\phi_+} \hskip -3pt d\phi''\Big\{
\frac{\rho^{+2}_\oplus}{2}\Big(J^2_0\big(\alpha\eta\rho^+_\oplus\big)+J^2_1\big(\alpha\eta\rho^+_\oplus\big)\Big)-
\frac{\rho^{-2}_\oplus}{2}\Big(J^2_0\big(\alpha\eta\rho^-_\oplus\big)+J^2_1\big(\alpha\eta\rho^-_\oplus\big)\Big)\Big\}.
\label{eq:power-blur5-out}
\end{eqnarray}

Similarly to (\ref{eq:power-int}), we present the integrand of (\ref{eq:power-blur5-out}) as
{}
\begin{eqnarray}
\frac{\rho^{+2}_\oplus}{2}\Big(J^2_0\big(\alpha\eta\rho^+_\oplus\big)+J^2_1\big(\alpha\eta\rho^+_\oplus\big)\Big)-
\frac{\rho^{-2}_\oplus}{2}\Big(J^2_0\big(\alpha\eta\rho^-_\oplus \big)+J^2_1\big(\alpha\eta\rho^-_\oplus\big)\Big)&=&\frac{\rho^+_\oplus(\phi'')-\rho^-_\oplus(\phi'')}{\pi\alpha\eta}.
\label{eq:power-int*}
\end{eqnarray}

With this results and using (\ref{eq:rho++}), expression (\ref{eq:power-blur5-out}) is evaluated as
{}
\begin{eqnarray}
P_{\tt off}({\vec x}_0)
&=&B_s \pi ({\textstyle\frac{1}{2}}d)^2 \frac{\mu_0}{z_0^2}
\frac{2}{\pi\alpha\eta^2}
\int_{\phi_-}^{\phi_+} \hskip -3pt d\phi''\sqrt{r_\oplus^2-{r_0}^2\sin^2\phi''}.
\label{eq:power-blur*33*}
\end{eqnarray}

Similarly to (\ref{eq:power-blur6*}), the result (\ref{eq:power-blur*33*}) indicates that, for a source with uniform surface brightness, there is no azimuthal dependence in power received from an off-image telescope position.

Remembering the definitions for $\mu_0$ from (\ref{eq:S_z*6z-mu2}), for $D$ from (\ref{eq:Dd}), for $\eta$ from (\ref{eq:power_rec2*}), and for $\alpha$ from (\ref{eq:power_rec2i}), we express (\ref{eq:power-blur*33*}) to the order of ${\cal O}\big({\overline z}{/z_0}\big)$ as
{}
\begin{eqnarray}
P_{\tt off}(r_0)
&=&B_s \pi^2 d^2 \frac{R_\oplus}{2z_0}\sqrt{\frac{2r_g}{\overline z}} \beta(r_0), \qquad r_0\geq r_\oplus,
\label{eq:power-blur*3*}
\end{eqnarray}
with the factor $\beta (r_0)$ given by the following expression:
{}
\begin{eqnarray}
\beta (r_0)
&=&\frac{1}{\pi}\int_{\phi_-}^{\phi_+} \hskip -3pt d\phi''\sqrt{1-\Big(\frac{r_0}{r_\oplus}\Big)^2\sin^2\phi''}=\frac{2}{\pi}{\tt E}\Big[\arcsin \frac{r_\oplus}{r_0},\Big(\frac{r_0}{r_\oplus}\Big)^2\Big],
\label{eq:beta_r0}
\end{eqnarray}
where ${\tt E}[a,x]$ is the incomplete elliptic integral \cite{Abramovitz-Stegun:1965}. The behavior of $\beta (r_0)$ is shown in the center plot in Fig.~\ref{fig:factors}. The combined behavior of $\epsilon(r_0)$ and $\beta(r_0)$ in the relevant range of distances is shown in the right-hand side plot of the same figure.

Equation (\ref{eq:power-blur*3*}) describes the light from the source that is present in the image plane outside the direct image of the exoplanet.  The existence of this signal is due to the specific optical properties of the SGL given by its PSF (\ref{eq:psf*}) which, as a function of the distance to the optical axis on the image plane, falls out much more slowly than the PSF of a regular telescope (\ref{eq:psf-norm}). This fact provides valuable insight for image recovery and the relevant work on prospective mission planning and development \cite{Turyshev-etal:2018,Turyshev-etal:2019-Decadal}.

\section{Amplification and angular resolution}
\label{sec:ampl-res}

If the receiving telescope's aperture is small, comparable in size to the width of the central peak of the SGL's PSF  (\ref{eq:S_z*6z-mu2}), the resulting observations are conducted in the wave optical regime, where the SGL possesses remarkable optical properties.  In this case, the SGL's magnification and its diffraction pattern are given as
\begin{eqnarray}
{ \mu}_{\tt SGL}^0&=&\frac{4\pi^2}{1-e^{-4\pi^2 r_g/\lambda}}\frac{r_g}{\lambda}\, J^2_0\Big(2\pi\frac{\rho}{\lambda}\sqrt{\frac{2r_g}{\overline z}}\Big)=1.12\times 10^{11}\,J^2_0\Big(48.98 \Big(\frac{\rho}{1\,{\rm m}}\Big)\Big(\frac{1\,\mu{\rm m}}{\lambda}\Big)\Big(\frac{650\,\textrm{AU}}{\overline z}\Big)^\frac{1}{2}\Big).
\label{eq:mu0}
\end{eqnarray}
The angular resolution in this case is determined from the size of that largest peak of (\ref{eq:mu0}) \cite{Turyshev-Toth:2017}:
\begin{eqnarray}
\delta\theta^0_{\tt SGL}&=&0.38\frac{\lambda}{\sqrt{2r_g \overline z}}
=0.10 \Big(\frac{\lambda}{1\,\mu{\rm m}}\Big)\Big(\frac{650\,\textrm{AU}}{\overline z}\Big)^\frac{1}{2}\, {\rm nas}.
\label{eq:ang-res}
\end{eqnarray}

However, given the fact that the Sun must be blocked by a coronagraph, using a small telescope that lacks the angular resolution to resolve the Sun's disk from the distance of the SGL's focal region is unpractical. Instead, systems with $1~{\rm m}$-class apertures are required for this purpose.  Such a telescope averages many lobes of the diffraction pattern \cite{Turyshev-Toth:2017}. This averaging erases the wave optical behavior of the SGL.  Light amplification and angular resolution are determined by the geometry of the problem.  To demonstrate this we note that, in the absence of the SGL, the power received from an object is given by the following expression:
\begin{eqnarray}
P^0_{\tt exo}&=&B_s \pi ({\textstyle\frac{1}{2}}d)^2 \frac{\pi R^2_\oplus}{z^2_0}.
\label{eq:Pexo0}
\end{eqnarray}
To evaluate the amplification of the SGL when observing extended resolved sources, in (\ref{eq:Pexo}), we factor out $P^0_{\tt exo}$ given by  (\ref{eq:Pexo0}), and present (\ref{eq:Pexo}) as
\begin{eqnarray}
P_{\tt exo}(r_0)&=&P^0_{\tt exo}  {\cal A}_{\tt SGL}(r_0),
\label{eq:PexoAmp}
\end{eqnarray}
where ${\cal A}_{\tt SGL}(r_0)$ is the SGL's light amplification for extended resolved sources:
{}
\begin{eqnarray}
{\cal A}_{\tt SGL}(r_0)&=&\frac{2z_0}{R_\oplus}\sqrt{\frac{2r_g}{\overline z}}\epsilon(r_0)=2.26\times10^6\, \epsilon(r_0) \,\Big(\frac{z_0}{30~\textrm{pc}}\Big)\Big(\frac{650\,\textrm{AU}}{\overline z}\Big)^\frac{1}{2}.
\label{eq:ampl}
\end{eqnarray}

We note that the result is independent on the wavelength and is determined in full by the geometry of the problem, the size of the object and position of the telescope in the image plane.

Angular resolution in this case is also determined by geometric considerations and the procedure of image sampling. Clearly, the maximal resolution is achieved when we can sample the entire surface of the source, namely when the number of linear pixels across the surface is given as $N_0={2r_\oplus}/{d}$. In this case, we achieve the highest angular resolution, $\delta\theta_0$, given as
{}
\begin{eqnarray}
\delta\theta_0=\frac{2R_\oplus}{N_0}\frac{1}{z_0}\equiv\frac{d}{\overline z}=
2.12 \,\Big(\frac{d}{1\,{\rm m}}\Big)\Big(\frac{650\,\textrm{AU}}{\overline z}\Big)\, {\rm nas}.
\label{eq:resol0}
\end{eqnarray}
However, it is hard to achieve such a sampling and thus to obtain such a resolution. It is more realistic to consider that we will be able to sample the image with $N\leq N_0$ linear pixels. In this, more conservative case, the angular resolution, $\delta\theta_N$, is
{}
\begin{eqnarray}
\delta\theta_N=\frac{2R_\oplus}{N}\frac{1}{z_0}=
\frac{2.84}{N}\Big(\frac{30\,\textrm{pc}}{z_0}\Big)\, \mu{\rm as}.
\label{eq:resol}
\end{eqnarray}

Although the realistic light amplification factor of the SGL (\ref{eq:ampl}) and the angular resolution (\ref{eq:resol}) that are achievable using a meter-scale observing telescope in the SGL's focal region are smaller than the theoretical maxima calculated in the wave optical regime, the values are still very impressive.
These results provide realistic insight into the potential use of the SGL for imaging of faint distant objects, such as exoplanets.

\section{Discussion and Conclusions}
\label{sec:disc}

We studied the image formation process with the SGL and  analyzed the power of the EM field received on a photometric detector of an imaging telescope.

We were able to estimate the power received from a disk with a uniform  surface brightness. A telescope with a modest aperture, $d\ll r_\oplus$,  traversing the image plane, will receive signals with different quantities of blur.  The relevant expressions are given by (\ref{eq:Pexo}) and (\ref{eq:power-blur*3*}). These expressions can be combined to describe the power of the received signal as a function of the telescope's position on the image plane. For a uniform source brightness, the result depends only on the separation from the optical axis, $r_0$:
{}
\begin{eqnarray}
P(r_0)=B_s\pi^2 d^2 \frac{R_\oplus}{2z_0}\sqrt{\frac{2r_g}{\overline z}}\,\mu(r_0),
\qquad{\rm with}\qquad
\mu(r_0)&=&
 \Big\{ \begin{aligned}
\epsilon(r_0), \hskip 10pt 0\leq r_0\leq r_\oplus& \\
\beta(r_0), \hskip 30pt r_0\geq r_\oplus& \\
  \end{aligned},
  \label{eq:Pexo-total}
\end{eqnarray}
where $\epsilon(r_0)$ and $\beta(r_0)$ are given by (\ref{eq:eps_r0}) and (\ref{eq:beta_r0}), correspondingly.  The blur's contribution is captured by factor $\mu(r_0)$, which, outside the directly projected image of the exoplanet, falls-off is $\propto 1/r_0$, as expected from the PSF of the SGL.

\begin{figure}
\includegraphics[width=0.55\linewidth]{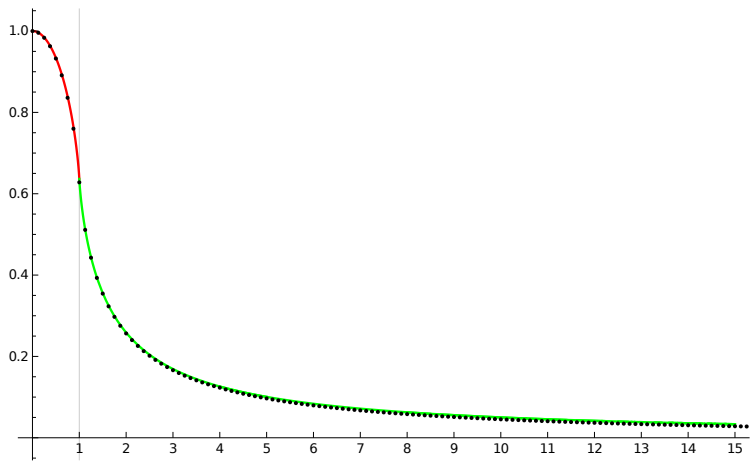}
\caption{\label{fig:validate}Blur analysis using numerically simulated and analytical results. Points represent the results of the numerical simulation shown in Fig.~\ref{fig:blur}, shown as a function of $r_0/r_\oplus$. The red curve is the analytical solution for the factor $\epsilon(r_0)$ from (\ref{eq:eps_r0}), while the green curve is thew same for the factor $\beta(r_0)$ from (\ref{eq:beta_r0}). Although, the results were obtained by two completely different methods, the plot shows nearly perfect agreement.}
\end{figure}

To validate the model (\ref{eq:Pexo-total}), we compared it against the blur contribution that corresponds to the numerically integrated result, shown in Fig.~\ref{fig:blur}.  The result of this analysis is shown in Fig.~\ref{fig:validate}. As we can see, our analytical model is in exact agreement with the results of the numerical simulation. As these two approaches address the problem in fundamentally different ways, we see the exact match shown in Fig.~\ref{fig:validate} as a validation of these two newly developed tools to study imaging with the SGL.

Fig.~\ref{fig:validate}  also highlights the fact that, because of the features of the PSF (\ref{eq:S_z*6z-mu2}), significant amounts of light from a faint target may be found in the off-image region. A failure to obtain light from this region can result in a loss of information, reducing the quality of reconstructed images and introducing noise artifacts. The impact of this behavior must be studied separately, taking into account the realistic SGL light amplification factor (\ref{eq:ampl}) and its relevant angular resolution (\ref{eq:resol}) that may be achieved when imaging extended sources with a modest size telescope.

Finally, the result (\ref{eq:Pexo-total}) may now be used to provide estimates for realistic signal levels that may be expected from various sources. As such, it will help to develop realistic signal-to-noise estimates that are needed for signal detection, processing, and image recovery. It may also be used for mission planning and development. The relevant work is a subject of an on-going study of imaging of exoplanets with the SGL. Results will be published as they become available.

\begin{acknowledgments}
This work in part was performed at the Jet Propulsion Laboratory, California Institute of Technology, under a contract with the National Aeronautics and Space Administration. VTT acknowledges the generous support of Plamen Vasilev and other Patreon patrons.

\end{acknowledgments}


\end{document}